\documentclass[submission,copyright,creativecommons]{eptcs}
\providecommand{\event}{THedu'11} 
\usepackage{breakurl}             
\usepackage{listings}
\usepackage{graphicx}
\usepackage{color}
\usepackage{wrapfig}

\usepackage[german, english]{babel}
\usepackage[T1]{fontenc}
\usepackage[latin1]{inputenc}

\long\def\symbolfootnote[#1]#2{%
  \begingroup%
     \def\thefootnote{\fnsymbol{footnote}}\footnote[#1]{#2}
  \endgroup}

\title{The GF Mathematics Library*}
\author{Jordi Saludes
\institute{Universitat Polit\`ecnica de Catalunya}
\institute{Sistemes Avan\c cats de Control}
\email{jordi.saludes@upc.edu}
\and
Sebastian Xamb\'o
\institute{Universitat Polit\`ecnica de Catalunya}
\institute{MA2, Edifici OMEGA, Barcelona (Spain)}
\email{\quad sebastia.xambo@upc.edu}
}
\def\titlerunning{MGL}
\def\authorrunning{J. Saludes \& S. Xamb\'o}
\begin{document}
\maketitle


\newcommand{\noun}[1]{\textsc{#1}}
\newcommand{\molto}{\textsc{mOlto}}
\newcommand{\webalt}{\textsc{WebALT}}
\newcommand{\openmath}{\textsc{OpenMath}}
\newcommand{\CD}{\textsc{CD}}
\newcommand{\OM}{\textsc{OM}}
\newcommand{\MGL}{\textsc{MGL}}
\newcommand{\MMA}{\textsc{MMA}}
\newcommand{\CAS}{\textsc{CAS}}
\newcommand{\CTP}{\textsc{CTP}}
\newcommand{\GF}{\textsc{GF}}
\newcommand{\Sage}{\textsc{Sage}}
\newcommand{\gfsage}{\texttt{gfsage}}
\newcommand{\Command}{\texttt{Command}}
\newcommand{\Answer}{\texttt{Answer}}

\newcommand{\Nat}{\texttt{Nat}}
\newcommand{\Prop}{\texttt{Prop}}

\newcommand{\exclude}[1]{}

\newcommand{\red}[1]{\textcolor{red}{#1}}
\newcommand{\blue}[1]{\textcolor{blue}{#1}}

\begin{abstract}
This paper is devoted to present the \emph{Mathematics
Grammar Library}, a system for multilingual mathematical text processing.
We explain the context in which it originated, its current design and
functionality and the current development goals.  We also present two
prototype services and comment on possible future applications in the area
of artificial mathematics assistants.%
\symbolfootnote[0]{%
*\ The research leading to these results has received funding
   from the European Union's Seventh Framework Programme (FP7/2007-2013)
   under grant agreement no. FP7-ICT-247914.}
\end{abstract}

\section{Introduction}
\label{intro}

An archetypal meeting point for natural language processing and mathematics
education is the realm of \emph{word problems}
\cite{Cooper-Harries-2005,Verschaffel-Greer-DeCorte-2000, Wayne-2001}, a
realm in which mechanised mathematics assistants (\MMA) are expected to
play an ever more prominent role in the years to come.

The following example, to which we will refer later on (last subsection
of \ref{sec:the_library}), is meant to illustrate in a concrete way the
idea of a word problem:
{\small
\begin{itemize}
\item[]
A farm has ducks and rabbits. There are $100$ animals and they have $260$
legs. How many ducks and rabbits are there in the farm?
\end{itemize}
}

We envision the Mathematics Grammar Library (\MGL) presented in this paper
as an enabling technology for \emph{multilingual} dialog systems capable of
helping students in solving and learning how to solve word problems.  This
confidence is grounded on the \MGL{} potential capabilities for dealing
effectively with a mixture of text and mathematical expressions,
capabilities that in turn depend crucially on the formal abstract way in
which the semantics is captured.

Since formal semantics is amenable to algorithmic processing, the library
can manage, in addition to parsing and rendering natural language with
mathematical expressions, powerful interactions with ancillary Computer
Algebra Systems (\CAS) or Computer Theorem Provers (\CTP{}).  As these are
key ingredients for advanced \MMA{}s, our working hypothesis is that \MGL{}
is a good basis on which to build useful \MMA's for learning and teaching
(cf.  \cite{E-LearningMathematics, AutonomousLearners} for some general
clues on e-learning technologies).

In this context, the current general aim for \MGL{} is to provide natural
language services for mathematical constructs at the level of high school
and college freshmen linear algebra and calculus. At the present stage,
the concrete goal is to provide rendering of simple mathematical exercises
in multiple languages (see \cite{Mathbar} for a demo of the expressions
available and also the examples in Section \ref{illustrations}).

For reading convenience, we include a short glossary of terms that will be
used in the rest of the paper.

\begin{description}
\item[\GF]
\href{http://www.grammaticalframework.org/}{Grammatical Framework}:
A programming language for multilingual grammar applications.
Based on functional programming and type theory, the framework supports
abstract grammars, which allow to capture meaning in a formal way,
and concrete grammars, which enable multilingual rendering.
See \cite{GF,Ranta11}. The library \MGL{} is programmed in \GF{},
in a way that is comparable to how numerical libraries are compiled from C
or Fortran sources.
\item[\openmath]
A \emph{de facto} standard for mathematical semantics, and usually
abbreviated as \noun{OM}. It is ``an extensible standard for representing
the semantics of mathematical objects, allowing them to be exchanged
between computer programs, stored in databases, or published on the
worldwide web'' (see \cite{OpenMath}).
It is structured in
\emph{Content Dictionaries} (\CD's), each of which
defines a collection of mathematical objects.
\item[\Sage]
Aimed at ``creating a viable free
open source alternative to Magma, Maple, Mathematica and Matlab'',
\Sage{} is the result of an on-going collective endeavour led
by William Stein. See
\cite{Kosan-2007, SageTutorial-2011} for a description of the system and
its functionalities.
\item[\webalt]
European digital content for the global networks project
(Contract Number EDC-22253). Developed in 2005 and 2006,
\href{http://webalt.math.helsinki.fi/content/}{\webalt}
aimed at using existing standards for representing mathematics on the web
and existing linguistic technologies to produce
language-independent mathematical didactical material.
See \cite{Caprotti06webalt!deliver, Caprotti_multilingualdelivery}.
\end{description}

\section{Background}
\label{sec:background}

For a closer view of \MGL, let us look briefly at its origins.  The
idea behind \MGL{} was born, to a good extend, on reflecting about one of
the key results of the \webalt{} project. In
summary, the unfolding of this reflection went as follows.

One of the aims of \webalt{} was to produce a proof-of-concept
platform for the creation of a multilingual repository of \emph{simple}
mathematical problems with guaranteed quality of the (machine)
translations, in both linguistic and mathematical terms. The languages
envisioned were Catalan,
English,
Finnish,
French,
Italian and
Spanish.
Of these, Finnish, with its great complexities, could not be raised to
the same level of functionality as the others.

The WebALT prototype was successful and, as far as we know, that endeavour
brought about the first application of the \GF{} system
for the multilingual translation of simple mathematical questions.  The
powerful \GF{} scheme, based on the perfect interlocking of abstract and
concrete grammars, was found to be a very sound choice, but the solution
had several shortcomings that could not be addressed in that project.
For the present purposes, the following three were the most appealing:
\begin{itemize}
\item
The grammars did not work for later versions of \GF{} ($>$2.9).
\item
The library was not modular with respect to
semantic processing, and hence not easy to maintain.
\item
It included too few languages, especially as seen from an European
perspective.
\end{itemize}

The springboard for the present library was the need to properly solve
these problems, inasmuch as this was regarded as one of the most promising
prerequisites for all further advanced developments in machine processing
of mathematical texts.  Thus the main tasks were:
\begin{itemize}
\item
To design a \emph{modular} mathematics library structured according to
the semantic standards (content dictionaries) of \openmath{}.
\item
To code it in the much more expressive \GF{}\,\,3.1 for the few languages
mentioned above, and
\item
To write new code for a few additional languages (Bulgarian, Finnish,
German, Romanian and Swedish).
\end{itemize}
The first two points amount to a tidying of the original \webalt{}
programming methods. The third point represents not merely an addition
of a few more languages, but a thourough testing of the methods and
procedures enforced in the preceding steps. This testing is important in
order to secure the rules for the inclusion of further languages and for
a controlled uniform extension of the available grammars.

To end this section, we include a few notions about the \GF{}
system that will ease the considerations about \MGL{} in the next section.
For a thourough reference about \GF, see \cite{Ranta11}.

Any GF application begins by specifying its \emph{abstract syntax}. This
syntax contains declarations of \emph{categories} (the \GF{} name for
types) and \emph{functions} (the \GF{} name for constructor signatures) and
has to capture the \emph{semantic structure} of the application domain.
For example, to let \Nat{} stand for the type of natural numbers and
\Prop{} for propositions about natural numbers, the \GF{} syntax is
{\small
\begin{verbatim}
     cat Nat, Prop ;
\end{verbatim}
}
That `zero is a natural number' and that `the successor of
any natural number is a natural number' can be expressed as follows:
{\small
\begin{verbatim}
     fun
        Zero : Nat ;
        Succ : Nat -> Nat ;
\end{verbatim}
}
The signatures for `even number' and `prime number' can be captured with
{\small
\begin{verbatim}
     fun
        Even, Prime : Nat -> Prop ;
\end{verbatim}
}
Finally, we can abstract the logical `not', `and' and `or' as follows:
{\small
\begin{verbatim}
     fun
        Not : Prop -> Prop ;
        And, Or : Prop -> Prop -> Prop ;
\end{verbatim}
}
In practical terms, these declarations would form the \emph{body} of an
\emph{abstract module} that would have the form
{\small
\begin{verbatim}
     abstract Arith = {<body>}
\end{verbatim}
}
\noindent where \texttt{Arith} is the name of the module.

\section{The \MGL{} library}
\label{sec:the_library}

As in any application coded in \GF{}, we need to specify what categories
will be used. In the case of \MGL{}, the most relevant categories
are in correspondence with all possible combinations of \textbf{Variable}
and \textbf{Value} with the mathematical types
\textbf{Number}, \textbf{Set}, \textbf{Tensor} and \textbf{Function}.
Thus the category \texttt{VarNum} denotes a numeric
variable like $x$, while \texttt{ValSet} denotes an actual set like ``the
domain of the natural logarithm''.  The distinction between variables and
values allows us to type-check productions like lambda abstractions
that require a variable as the first argument. Variables can be
promoted to values when needed.

The library is organised in a matrix-like form, with an horizontal axis
ranging over the targeted natural languages.  At the moment these are:
Bulgarian,
Catalan,
English,
Finnish,
French,
German,
Italian,
Polish,
Romanian,
Spanish,
Swedish
and
Urdu.
In addition, the mathematical typesetting system \LaTeX{} has also been
included, and also a natural language interface to \Sage{}
that allows to elicit results from this sophisticated
computational environment with commands expressed in any of the natural
languages currently available.

The vertical axis is for complexity and contains, from bottom to top, three
layers:

\begin{description}
\item[\emph{Ground}.] It deals with literals, indices and variables.
\item[\emph{OpenMath.}]
It is modelled after the \OM{ } \emph{Content
Dictionaries} (\CD's), in the sense that in this layer there is an \MGL{}
module for each \CD.
\item[\emph{Operations.}]
This layer takes care of simple mathematical exercises. These appear
in drilling materials and usually begin with directives such as
`Compute', `Find', `Prove', `Give an example of', \ldots.
\end{description}

The following tree is an example of what can be expressed in the
\emph{OpenMath} layer:
{\small
\begin{verbatim}
     mkProp
     (lt_num
       (abs (plus (BaseValNum (Var2Num x) (Var2Num y))))
       (plus (BaseValNum (abs (Var2Num x)) (abs (Var2Num y)))))
\end{verbatim}
}
When linearized, say with the Spanish concrete grammar, it yields
{\small
\begin{quote}
El valor absoluto de la suma  de x y de y es menor que la suma del valor
absoluto  de x y del valor absoluto de y
\end{quote}
}

Similarly, the tree
{\small
\begin{verbatim}
     DoSelectFromN
       (Var2Num y)
       (domain (inverse tanh))
       (mkProp
         (gt_num
         (At cosh (Var2Num y))
          pi))
\end{verbatim}
}
gives, when linearized with the English concrete grammar:
{\small
\begin{quote}
Select y from the domain of the inverse of the hyperbolic tangent such that
the hyperbolic cosine of y is greater than pi.
\end{quote}
}

\label{illustrations}
We end this section by describing two prototype services driven by \MGL{}:
the Mathbar demo and the \gfsage{} service.

\subsubsection*{Mathbar demo}

To access this demo, see \cite{Mathbar}. Now consider, for example, the
sentence ``Gamma is greater than pi raised to
x'', which can be easily composed by choosing Eng in the \texttt{From} slot
and repeatedly choosing the desired word among the continuation options
presented at each stage. If we further choose All in the \texttt{To} slot,
we get the results shown in the screenshot.


\begin{figure}[ht]
\begin{center}
\includegraphics[width=110mm]{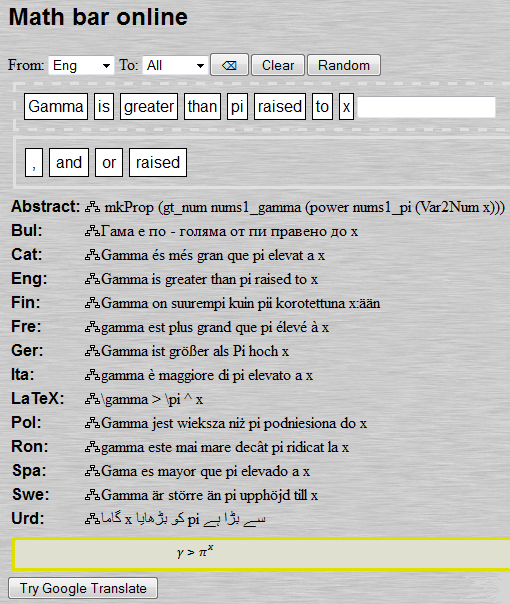}
\end{center}
\end{figure}

%

At the bottom, we can see the typesetting of \LaTeX{} of the expression
``$\backslash$gamma $>\backslash$pi\^{} x'':
\[
\gamma>\pi^x
\]

\paragraph{Remark.}
There are a few details in some of the concrete grammars that have to be
improved.
In the case of Polish, `podniesiona' should be
`podniesione', because `pi' is neutral in that language, and `wieksza'
should be `wi\c eksza' (Adam Slaski, private communication).
There is also a slight inconsistency in the rendering of `Gamma', since
in French, Italian and Romanian it appears with `g' while for
all other it goes with `G'. Actually it is not hard to modify the
linearizations so that they produce
`$\pi$', `$\gamma$' and `$\Gamma$' instead of `pi', `gamma' and `Gamma'.

\paragraph{Remark.}
In the Mathbar demo there is the button ``Try Google Translate''. When we
try for the different languages, there are cases in which we get the same
result (Catalan, Romanian, Spanish, Swedish), but in others
the result is different, and often wrong:

\begin{center}
\begin{tabular}{ll}
\label{mathinfredge}
Bul & \includegraphics[width=65mm]{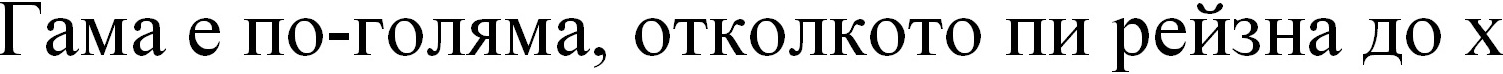} \\[-1pt]
Fin & Gamma on suurempi kuin PI nostetaan x\\
Fre & Gamma est sup\'erieure \`a Pi port\'ee \`a x\\
Ger & Gamma gr\"oßer als pi um x erh\"oht\\
Ita & Gamma è superiore a pi elevato a x \\
Pol & Gamma jest wi\c eksza ni$\dot{\textrm{z}}$ pi podniesiony do x\\
Urd & \includegraphics[width=35mm]{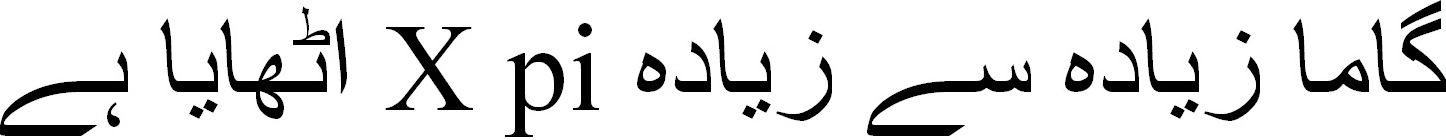}
\end{tabular}
\end{center}

\subsubsection*{\Sage{} commands in natural language}


Another recently developed prototype based on \MGL{} is \gfsage.  It
enables to express \Sage{} commands in natural language and get the results
expressed likewise.  The tool starts a \Sage{} notebook server in the
background (as described in Simple Sage Server API, \cite{SSSAPI}), reads
the pgf grammar file and translates the queries from the chosen natural
language to the concrete grammar for \Sage.  This is passed to the \Sage{}
server for evaluation.  Each computation runs in a different worksheet cell
and the server replies with a \texttt{done} or a \texttt{computing}
message.  In this case the program waits for completion of the computation
and then writes the answer.

From the GF side, what is send to \Sage{} is always in the category
\Command.  What is returned by \Sage{} is in the category \Answer.  There
are 3 kinds of Commands:

\begin{itemize}
\item
    Asking for a computation.
    Compute: \texttt{Kind -> Value Kind -> Command}.\par
    \Sage{} gives back a \texttt{ReturnBlock} with the cell number and the
    answer (a string). We could now construct a short \Answer{} by using:
\begin{itemize}
\item
      Simple: \texttt{k $\in$ Kind -> Value k -> Answer}\\
      (``it is 5''), or
\item
      Feedback: \texttt{k $\in$ Kind -> Value k -> Value k -> Answer}\\
      (``the factorial of 3 is 6''), that combines the question (the first
       Value k) with the \Sage{} answer.
\end{itemize}
\item
    Assuming propositions.
    Assume: \texttt{Prop -> Command}.\par
    \Sage{} silently accepts the command by returning an
    \texttt{EmptyBlock} (with cell number) but we want it to be more
    assertive, so we reinject the
    \texttt{Prop} into Assumed: \texttt{Prop -> Answer}\\
    (``I assume that x is greater than 2'')
\item
    Binding Values to Variables. \par
    Assign:
    \texttt{k $\in$ Kind -> Var k -> Value k -> Command}\\
    (``assign 2 to x''). \par
    We expect \Sage{} to return an \texttt{EmptyBlock} followed by\par
    Assigned:
   \texttt{k $\in$ Kind -> Var k -> Value k -> Answer}\\
   (``2 is now assigned to x'').
\end{itemize}

Here are some illustrations:

{\small
\begin{verbatim}
     sage> compute the sum of 1, 2, 3, 4 and 5.
     [4] 15
     answer: it is 15
\end{verbatim}
}

{\small
\begin{verbatim}
     sage> compute the summation of x when x ranges from 1 to 100.
     [7] 5050
     answer: it is 5050
\end{verbatim}
}

{\small
\begin{verbatim}
     sage> compute the integral of the cosine on the open interval
           from 0 to the quotient of pi and 2.
     [8] 1
     answer: it is 1

     sage> compute the integral of the function mapping x
           to the square root of x on the closed interval from 1 to 2.
     waiting...
     [4] 4/3*sqrt(2) - 2/3
     answer: it is 4/3*sqrt(2) - 2/3
\end{verbatim}
}

{\small
\begin{verbatim}
     sage> compute the sum of x and y.
     [4] x + y
     answer: it is x plus y.

     sage> compute the sum of x and 5.
     [5] x + 5
     answer: it is x plus 5.

     sage> compute the sum of 4 and 5.
     [6] 9
     answer: it is 9.
\end{verbatim}
}

\newenvironment{itemize*}%
  {\begin{itemize}%
    \setlength{\itemsep}{0pt}%
    \setlength{\parskip}{0pt}}%
  {\end{itemize}}

\newenvironment{enumerate*}%
  {\begin{enumerate}%
    \setlength{\itemsep}{0pt}%
    \setlength{\parskip}{0pt}}%
  {\end{enumerate}}

\subsubsection*{Dealing with word problems}

Let us return to our word problem example in Section \ref{intro} in order
to consider the difficulties posed by a full computer representation of its
more relevant aspects, and also to point out some hints about how to
achieve it. First of all, there is the question that
human readers are expected to make sense of information that is not stated
explicitely but which they usually infer from the semantic context.  In our
example, it is enough to write the inferred assertions next to the
assertions given in the word problem.  Notice that some of the inferences
amount to making explicit the implicit references.

\vspace{1ex}

\begin{tabular}{ll}
A farm has ducks and rabbits. &
  1. A farm has \emph{no animals other than} \\
  & \quad ducks and rabbits. \\[3pt]
There are $100$ animals &
    2. There are 100 animals \emph{in the farm}. \\[3pt]
and they have $260$ legs. &
    3. The animals \emph{in the farm} have 260 legs. \\[3pt]
How many ducks and rabbits are there in the farm? &
    4. How many ducks are there in the farm?\\[3pt]
   & \quad  How many rabbits are there in the farm?
\end{tabular}

\vspace{1ex}
Let us proceed now with a few
hints about how the right-hand side statements in the table
could be elicited from the left-hand side ones.

\begin{enumerate*}
\item The line can be parsed except for \textbf{farm}, \textbf{ducks} and
\textbf{rabbits}, which are unknown to \MGL.
It can be inferred, however (using the structure available from the \GF{}
parser), that these unknowns are \emph{common nouns}. Then
a query to Wordnet \cite{wordnet} finds entries compatible with this
assumption. From the determinants used, we deduce that there is an
instance $f$ of the entity \noun{farm}
($F$) and that there are entities \noun{duck} and \noun{rabbit} ($D$
and $R$, respectively). The verb \textbf{has} is a priori related to
the \noun{in} predicate:%
\symbolfootnote[0]{%
*\ It is not hard to have instances of the
problem whose solution has no rabbits (or no ducks).
This may come as a surprise to the student, but it is mathematically
acceptable by common practice.
If we were to follow this convention, then
we would drop these inequalities.}
\[
f\in F,\quad|D\cap\mbox{\textsc{in}}(f)|\geq1,\textrm{*}
\quad|R\cap\mbox{\textsc{in}}(f)|\geq1,\textrm{*}
\quad A\cap\mbox{\textsc{in}}(f)\setminus(D\cup R)=\emptyset.
\]
\item \textbf{Animals} is a new common noun leading to a new entity $A$.
Another query to Wordnet reveals that it is, in fact, an hypernym of \textbf{duck}
and \textbf{rabbit}:
\[
|A\cap\mbox{\textsc{in}}(f)|=100,\quad D,R\subset A.
\]

\item The noun \textbf{legs} gives rise to another entity ($L$) and the
occurrence of \textbf{have} introduces a new version of \noun{in}:
\[
|L\cap\mbox{\textsc{in}}(A\cap\mbox{\textsc{in}}(f))|=260.
\]

\item Wordnet points out that a \textbf{farm} is a \emph{location},
so \textbf{there} probably refers to $f$. A \textbf{how many} question
asks for $d=|D\cap\mbox{\textsc{in}}(f)|$ and $r=|R\cap\mbox{\textsc{in}}(f)|$.
\end{enumerate*}

\section{Conclusions and further work}

In this paper we have described a \GF{} library, which we call \MGL{},
for multilingual mathematical text processing.
We have also indicated how it
originated in the \webalt{} project, its relation to \GF{}, and its
present functionality.
After a first step in which the main concern was tidying and modularizing
the \webalt{} prototype for simple mathematics exercises in five languages,
we have extended it, in a second step, with four more languages
(Finnish was considered in the first step, but
it had to be worked out from scratch in the second step).
We have also showed that \LaTeX{} and \Sage{} can be approached with the
same methodology. In particular, \gfsage{} allows to interact with \Sage{}
by expressing the commands in natural language.

\vspace{2ex}
Further work has three main lines:
\begin{itemize}
\item
Addition of new languages,
like Danish, Dutch, Norwegian, Portuguese, Russian, \ldots
This is a continuation of the first two steps referred to above and
our assessment is that it can be done reliably with the methods and
procedures established so far. To some extent, the library modules for a
new language can be generated automatically up to a point in which the
remaining work corresponds to natives in that language.
\item
Describing a systematic procedure for the uniform and reliable extension
of the grammars according to new semantic needs. This is an important step
that is being researched from several angles. One important point is to
ascertain when a piece of mathematical text requires functionalities
(categories, constructors, operations) not yet covered by MGL.
\item
Advancing in the use of \MGL{} for the production of ever more
sophisticated artificial mathematics assistants. This is also the focus of
current research that includes a collaboration with statistical machine
translation methods, as in principle they can suggest grammatical structures
out of a corpus of mathematical sentences. One important element will be an
extended version of \gfsage{} that will enable to harness a powerful CAS
system such as \Sage{} by means of commands expressed in natural
languages. We also envision a similar prototype to harness the capabilities
of \CTP{}s. After this, we hope that we will be in a position to produce
a \MMA{} that can help students in solving and learning how to solve word
problems of the kind we have been considering.
\end{itemize}

\subsection*{How to get \MGL}

The living end of the library is publicly available using subversion as:
\begin{verbatim}
     svn co svn://molto-project.eu/mgl
\end{verbatim}
A stable version can be found at:
\begin{verbatim}
    svn co svn://molto-project.eu/tags/D6.1
\end{verbatim}

\subsection*{Acknowledgments}

The authors are much grateful to the referees for pointing out several ways
in which this paper could be improved, especially by insisting that we made
more explicit the features of our work that are closer to the conference
scope.  We are also thankful to Pedro Quaresma for his kind and helpful
handling of the correspondence concerning this paper.


\end{document}